\documentclass[structabstract]{aa}  
\usepackage{graphicx}
\usepackage{txfonts}
\usepackage{natbib}
\bibpunct{(}{)}{;}{a}{}{,}
\begin{document}
   \title{Detection of unidentified infrared bands in a H$\alpha$ filament
   in the dwarf galaxy NGC1569 with AKARI}


   \author{Takashi Onaka
          \inst{1}
          \and
          Hiroko Matsumoto\inst{1}\fnmsep\thanks{Present address:
          Nikkei Inc. Chiyoda-ku, Tokyo 100-8066, Japan}
          \and 
          Itsuki Sakon
          \inst{1}
          \and
          Hidehiro Kaneda
          \inst{2}
          }

   \institute{Department of Astronomy, Graduate School of Science,
              The University of Tokyo,
              Bunkyo-ku, Tokyo 113-0033, Japan\\
              \email{onaka@astron.s.u-tokyo.ac.jp; 
              isakon@astron.s.u-tokyo.ac.jp}
         \and
             Graduate School of Science, Nagoya University
             Chikusa-ku, Nagoya 464-8602, Japan\\
             \email{kaneda@u.phys.nagoya-u.ac.jp}
             }

   \date{}

 
  \abstract
   {We report the detection of unidentified infrared (UIR) bands in a 
   filamentary structure associated with H$\alpha$ emission in the
   starburst dwarf galaxy NGC1569 based on imaging and spectroscopic
   observations of the {{\it AKARI}} satellite.}
   {We investigate the processing and destruction of the UIR band carriers in an outflow
   from NGC1569.
   }
   {We performed observations of NGC1569 for 6 infrared bands (3.2, 4.1, 7, 11, 15,
   and 24\,$\mu$m) with the Infrared Camera
   (IRC) onboard {\it AKARI}. Near- to mid-infrared
   (2--13\,$\mu$m) spectroscopy of a H$\alpha$ filament was also carried out 
   with the IRC. }
   {The extended structure associated with a H$\alpha$ filament appears
   bright at 7\,$\mu$m.  Since the IRC 7\,$\mu$m band
   (S7) efficiently traces the 6.2 and 7.7\,$\mu$m UIR band emission, the 
   IRC imaging observations
   suggest that the filament is bright at the UIR
   band emission.  Follow-up spectroscopic observations with the IRC confirm
   the presence of 6.2, 7.7, and 11.3\,$\mu$m emission in the filament.   The
   filament spectrum exhibits strong 11.3\,$\mu$m UIR band emission relative 
   to the 7.7\,$\mu$m band compared to the galaxy disk observed with the
   Infrared Spectrograph on {\it Spitzer}.  The near-infrared spectrum also
   suggests the presence of excess continuum 
   emission in 2.5--5\,$\mu$m in the filament.
   }
   {The presence of the UIR bands associated with a H$\alpha$ filament
   is found by {\it AKARI}/IRC observations.  The H$\alpha$ filament is thought
   to have been formed by the galactic 
   outflow originating from the star-formation
   activity in the disk of NGC1569.
   The destruction timescale of the UIR band carriers in the outflow
   is estimated
   to be much shorter ($\sim 1.3 \times 10^3$\,yr) than the
   timescale of the outflow ($\sim 5.3$\,Myr).  Thus it is unlikely
   that the band carriers survive the outflow environment.
   Alternatively, we suggest that the band
   carriers in the filaments may be produced by the fragmentation of large
   carbonaceous grains in shocks, which produces the H$\alpha$ emission.
   The NIR excess continuum emission cannot be accounted for by free-free
   emission alone and a hot dust contribution may be needed,
   although the free-free emission intensity estimated from \ion{H}{i}
   recombination lines has a large uncertainty.}


   \titlerunning{Detection of the UIR bands in a NGC1569 filament}
   \authorrunning{T. Onaka et al.}   
   \maketitle

%

\section{Introduction}
From the near-infrared (NIR) to mid-infrared (MIR) region, 
a set of prominent emission features 
at 3.3, 3.4, 6.2, 7.7, 8.6, and 11.3\,$\mu$m are detected for the various
celestial objects.  
Faint companion bands are also present in the 5--13\,$\mu$m  spectral range.
They are often called the unidentified infrared (UIR) bands
because of the difficulty in identifying the band carriers at an 
early epoch of their discovery.
Space observations have added the 17\,$\mu$m feature complex to
the UIR band family \citep{kerckhoven00, werner04}.
The UIR bands have been observed in \ion{H}{ii} regions, reflection nebulae,
post-asymptotic giant branch (AGB) stars, and planetary nebulae (PNe)
\citep[e.g.,][]{peeters02}.
They are also 
ubiquitously seen in the diffuse Galactic radiation 
\citep{giard88, onaka96, mattila96, tanaka96, onaka04} as well as in external 
galaxies
\citep{helou00, lu03, smithjd07}, suggesting that the band carriers belong
to major constituents of the interstellar matter \citep[e.g.,][]{draine07}.
Since their intensity is found to be well correlated with the far-infrared
(FIR) intensity and can thus be used as a useful measure of the star-formation
activity \citep{onaka00, peeters04}, and they are prominent features
in the MIR that can be detected even in distant galaxies \citep[e.g.,][]{lutz05},
the understanding of the properties as well as 
the formation and destruction processes in the interstellar space of the band 
carriers is significant not only to the study of the interstellar medium
(ISM), but also to the study of physical conditions and star-formation
activities in the remote universe.

While it is generally thought that the UIR bands originate in  
emitters or emitting atomic groups containing polycyclic aromatic 
hydrocarbons (PAHs) or PAH-like atomic groups of carbonaceous materials
\citep{leger84, allamandola85, sakata84, papoular89},
the exact nature, formation, and destruction of the band carriers are not
yet fully understood \citep[see][for a recent review]{tielens08}.  
Carbon-rich AGB stars are thought to be 
one of the major sources of the band carriers \citep{frenklach89,
latter91, cherchneff92, galliano08a}.  
The carriers may also be formed by the fragmentation of 
large carbonaceous grains \citep{omont86, jones96, greenberg00} or in situ
within dense clouds \citep{herbst91}.
On the other hand, the carriers can 
be destroyed efficiently by interstellar shocks
\citep{jones96} and within ionized gas 
\citep[e.g.,][references therein]{matsumoto08}.
Few observational studies have, however, so far been carried out for 
the life cycle of the band carriers in the ISM.

\citet{peeters02} show that there are at least three distinct classes of UIR-band objects
present as far as the peak wavelengths of the 6.2, 7.7, and 8.6\,$\mu$m MIR
UIR bands are concerned.
Class A objects, to which \ion{H}{ii} regions and isolated Herbig Ae/Be
stars belong, show the band peaks at short wavelengths relative
to Class B objects, which include PNe and post-AGB stars.  Class C objects
exhibit quite different spectra compared to class A and B objects.
The variation in the peak wavelengths suggests possible processing of the
band carriers in the ISM.
The origin of the difference between class A and B objects may be attributed
to the inclusion of hetero atoms \citep{peeters02} or carbon isotopes
in the band carriers \citep{wada03}.  On the other hand, the UIR band spectrum
observed in the diffuse Galactic radiation does not exhibit any
systematic variations in the inner part of the Galaxy
\citep{chan01, kahanpaa03}, whereas \citet{sakon04} detected small variations
in the band ratio and the peak wavelength
between the inner part of the Galaxy and the outer part.  Variations
in the band spectra of external galaxies and within a given galaxy
are also generally small \citep[e.g.,][]{smithjd07, gordon08}.  
Studies, however,
indicate that elliptical galaxies show UIR band spectra very different 
from those in spiral galaxies,
where the usually strong MIR UIR bands (6.2, 7.7, and 8.6\,$\mu$m)
are weak or absent, but the 11.3\,$\mu$m
band is clearly present \citep{kaneda05}.  The 3.3\,$\mu$m band seems also to be
absent in the elliptical galaxy NGC1316, which exhibits the 11.3\,$\mu$m band
\citep{kaneda07}.  The weak MIR UIR bands relative to that at 11.3\,$\mu$m are
also seen in galaxies that harbor a low-luminosity active galactic nucleus 
\citep[AGN,][]{smithjd07}, some of which are also elliptical galaxies.
Smaller scale variations in the same sense have also been
detected in the interarm region of NGC6946 \citep{sakon07} and
halo regions of galaxies \citep{irwin06, galliano08b}, suggesting that a common
mechanism changes the band ratio \citep{onaka08}.  
Weak MIR UIR bands may be seen in tenuous plasma environments.  
Processing of the band carriers
in these environments may be responsible for the variation.

The UIR bands in low-metallicity dwarf galaxies are important to study because
their characteristics may represent very young galaxies and the band carriers
may not be fully developed in these environments if carbon-rich AGB stars are the
main source of the carriers \citep{galliano08b}.  The UIR bands are very weak
or almost absent in dwarf galaxies that have metallicities log[O/H]+12 $\la$ 8.1 
\citep{engelbracht08}.  Since these low-metallicity dwarf galaxies are also known
to be associated with high star-formation activities,
the deficiency in the UIR bands in low metallicity environments can be
attributed to low carrier formation efficiency at low metallicity, rapid destruction
of the band carriers by strong radiation fields \citep{wu06, wu07}, 
frequent supernova shock passages that destroy the carriers \citep{ohalloran06}, 
or deficiency in carbon-rich AGB stars that produce the carriers
in young galaxies \citep{galliano08b}.

In this paper, we report the results of infrared imaging and spectroscopic
observations of the low-metallicity dwarf galaxy \object{NGC1569} with the Infrared
Camera (IRC) onboard {\it AKARI} \citep{murakami07, onaka07b}.
NGC1569 is a nearby starburst dwarf galaxy of metallicity of about
a quarter of solar, i.e., its values of 12 + log(O/H) range from 8.19 to 8.37 \citep[]
[references therein]
{greggio98}.  {\it Hubble Space Telescope} observations
indicate that it is located  in the IC 342 group of
galaxies at a distance of
$3.36\pm0.20$\,Mpc \citep{grocholski08}. Several filamentary structures that 
extend
out to 1\,kpc have been detected in H$\alpha$ \citep{hunter93, heckman95,
martin98, west07a, west07b, west08}.  
Associated X-ray emission has also been
detected, suggesting that these filaments are formed by the galactic
outflow \citep{martin02}.  NGC1569 has two well-known super-star clusters
(SSCs) and numerous compact clusters \citep{hunter00}.  Some remain
deeply embedded in dense clouds \citep{tokura06}.  NGC1569 has probably
experienced three major epochs of star formation over the entire
galaxy that peaked at 5--27, 32--100, and $\ga$2000\,Myr ago \citep{greggio98,
angeretti05, grocholski08}.  Shock-ionized gas appears to be detected 
\citep
{buckalew00, buckalew06}, which is indicative of recent star-formation activity
\citep{west07b}.  The 
\ion{H}{i} inflow stream may have interacted with the galaxy disk
in the region of a velocity-crowding 'hot spot' 
and triggered the star formation \citep{stil98, muhle05}.

The metallicity of NGC1569 is slightly above the threshold for the 
presence of UIR bands.
ISOCAM observations of NGC1569 show that the entire galaxy disk is
dominated by UIR band emission \citep{madden06}, although it is weak
or absent locally around SSCs and \ion{H}{ii} regions \citep{tokura06}.
To investigate the distribution and possible spatial variations in the UIR
bands in NGC1569, imaging and spectroscopic observations were executed
with {\it AKARI} as part of the mission program  
``ISM in our Galaxy and Nearby Galaxies'' \citep[ISMGN:][]{kaneda09a}.
The observations and the data reduction are described in Sect. 2 and the
results are presented in Sect. 3.  The origin of the UIR band carriers
is discussed in Sect. 4.  A summary is given in Sect. 5.


\section{Observations and data reduction}
IRC imaging observations of NGC1569 were carried out on 2006 September 9
in the two-filter mode \citep[AOT: IRC02,][]{onaka07b}.  The size
of the galaxy is a few arcminutes.  Owing to the wide field-of-view
of the IRC ($\sim 10^\prime \times 10^\prime$), one pointing observation
covers the entire galaxy. 
With two pointing observations, we obtained images in the 6 bands
N3 (3.2\,$\mu$m), N4 (4.1\,$\mu$m), S7 (7.0\,$\mu$m), S11 (11.0\,$\mu$m),
L15 (15.0\,$\mu$m), and L24 (24.0\,$\mu$m) because
the NIR (N3 and N4) and the MIR-S (S7 and S11) channels share the
field-of-view and observe the same area of sky at the same time.
The MIR-L (L15 and L24) observes a sky about $20\arcmin$ from the
NIR/MIR-S.  

Spectroscopic observations were carried out with AOT IRC04 with the
NIR and MIR-S slit mode (Ns), whose width is 5\arcsec and length is about
1\arcmin, for 1.8--13.4\,$\mu$m.  The center of the slit was 
placed at the position of a
H$\alpha$ filament, where the presence of the UIR bands was
inferred from the imaging observations (see Table 2 and Fig.~\ref{fig:color}b).  
The spectrum consists of 3 segments.
The spectrum of 1.8--5.5\,$\mu$m was taken with the prism (NP) in the NIR
channel, while the spectra of 4.6--9.2\,$\mu$m and
7.2--13.4\,$\mu$m were taken with two grisms, SG1 and SG2, in the MIR-S, 
respectively.  
Table~\ref{tab:obs} summarizes the observation log.

   \begin{table*}[!ht]
      \caption[]{IRC observation log of NGC1569}
         \label{tab:obs} 
       \centering
      \begin{tabular}{l l l l l}
      \hline
      \hline
 	Obs. mode & Obs. ID & Date & AOT$^{\mathrm{a}}$ & Filter \\
      \hline
 	Imaging & 1400423.1 & 2006 September 9 & IRC02 a:N & N3, N4, S7, \& S11\\
 	        & 1400424.1 & 2006 September 9 & IRC02 a:L & L15 \& L24 \\
 	Spectroscopy & 1400701.1 & 2007 March 8 & IRC04 a:Ns & 
	prism (NP) \& grisms (SG1 and SG2)\\
      \hline
      \end{tabular}
      \begin{list}{}{}
      \item[$^{\mathrm{a}}$]Astronomical Observation Template.  See
      ASTRO-F Observer's Manual 
      for details of the parameters 
      (http://www.ir.isas.jaxa.jp/ASTRO-F/Observation/ObsMan/afobsman32.pdf).
      \end{list}
   \end{table*}

The imaging data were processed with the IRC imaging toolkit version
20071017, which includes linearity correction, dark current
subtraction, flat-fielding, correction for the image distortion,
and coaddition of dithered images.  The toolkit algorithm produces final
images with the pixel field-of-view of
$1\farcs446$ for N3 and N4, $2\farcs340$ for
S7 and S11, and $2\farcs384$ for L15 and L24.
The images produced by the toolkit were processed by our own
software.  The original images are in the array coordinates.
An area of $5\farcm5 \times 5\farcm0$ is extracted
in the equatorial coordinates and then smoothed with a Gaussian
of FWHM $3\farcs5$ for the N3 and N4 images
and $7\arcsec$ for the other 4 bands to adjust the spatial resolution
with respect to each other without appreciable degradation of the image quality in each 
band \citep{onaka07b}.  
The contribution from the
stellar photospheric emission is subtracted except for the N3 image, 
which is used as a reference after subtracting the constant sky background.
The flux ratio relative to N3 in units of ADU is 
estimated for the average stellar spectrum based on
the fluxes of red giant stars used in the flux calibration as
0.69, 0.073, 0.050, 0.0098, and 0.0017 for N4, S7, S11, L15, and
L24, respectively \citep{tanabe08}.
The stellar flux subtraction is important only in the galactic disk region
and does not make a significant effect except for the N4 image.

The flux calibration of the IRC for point sources 
is more accurate than 10\% for all the photometric bands
\citep{tanabe08}, though the calibration for the diffuse emission is still underway.
The uncertainty of about 30\% is expected for the absolute calibration for the
diffuse emission because
the IRC uses the same types of
detector arrays as the IRAC on {\it Spitzer} \citep{reach05, cohen07}.
Since the relative flux has a far smaller uncertainty, the following discussion focuses
on the relative fluxes among the bands or colors
rather than on the absolute flux values.

The spectroscopic data were processed with the IRC spectroscopy toolkit 
version 20090211.  Each segment of the spectrum was truncated at the wavelengths
where the signal-to-noise ratio becomes low: NP was truncated at 2.3 and
5.0\,$\mu$m, SG1 at 5.5 and 8.2\,$\mu$m, and SG2 at 8.0 and 13.2\,$\mu$m.
The accurate position at which
the spectrum was taken was estimated for the reference image of N3 obtained during 
the same pointing observation.   
The original spectrum has contributions from NGC1569, the diffuse Galactic 
emission,
and the zodiacal light.  Since we did not perform separate observations to obtain 
the background spectrum, we chose a "sky" on the slit near the edge, where
the contribution of NGC1569 is minimal, 
and subtracted the sky spectrum from the "filament" spectrum.  
Table~\ref{tab:spectrum} summarizes the center positions where the spectra 
of the filament and sky were extracted (see also Fig.~\ref{fig:color}b).
The pixel scale was different between the NIR and MIR-S channels.  
The spectra were extracted for a slit length of 
5 and 3 pixels ($7\farcs3$ and $7\farcs0$) for the NIR and MIR-S,
respectively, and averaged.  We note that even for the SG1 and SG2 spectra
the observed position was slightly different because the SG1 and SG2 grisms
did not produce the same optical path.  
The SG1 spectra were smoothed by taking a moving average over 4 pixels in
the spectral direction
to increase the signal-to-noise ratio without significant degradation
of the spectral resolution.  The SG2 spectra were smoothed
by a 3-pixel moving average.  The NP spectrum of the filament longer than
4.3\,$\mu$m was smoothed by taking a 2-pixel running average, while
that shortward of 4.3\,$\mu$m was not smoothed to prevent decreasing the
spectral resolution since NP spectra have lower resolution at shorter
wavelengths.  These smoothing processes produced a spectral resolution 
($\lambda / \Delta\lambda$) of about 30 at the central wavelength of each segment.
The raw sky spectrum of NP had  a low signal-to-noise ratio and was 
smoothed by a 5-pixel moving average in the wavelength direction based on the assumption
that the sky spectrum did not have significant features.

   \begin{table}[!ht]
      \caption[]{Positions of the spectrum extracted}
         \label{tab:spectrum} 
       \centering
      \begin{tabular}{l c c }
      \hline
      \hline
 	Object & \multicolumn{2}{c}{Center position (J2000)$^{\mathrm{a}}$}  \\
      \hline
                      & R.A. & Dec \\
 	Filament & 04$^\mathrm{h}$ 30$^\mathrm{m}$ 41$\fs$1 &  64\degr 51\arcmin 03\farcs4 \\
 	Sky & 04$^\mathrm{h}$ 30$^\mathrm{m}$ 39$\fs$0 &  64\degr 50\arcmin 59\farcs8\\
      \hline
      \end{tabular}
      \begin{list}{}{}
      \item[$^{\mathrm{a}}$] The spectrum was extracted over $5\arcsec \times 7\farcs3$ and
      $ 5\arcsec \times 7\farcs0$ regions, for the NIR and MIR-S, respectively (see text).
      \end{list}
   \end{table}

\section{Results}
\subsection{Imaging}
\begin{figure*}
   \centering
   \includegraphics[width=\textwidth]{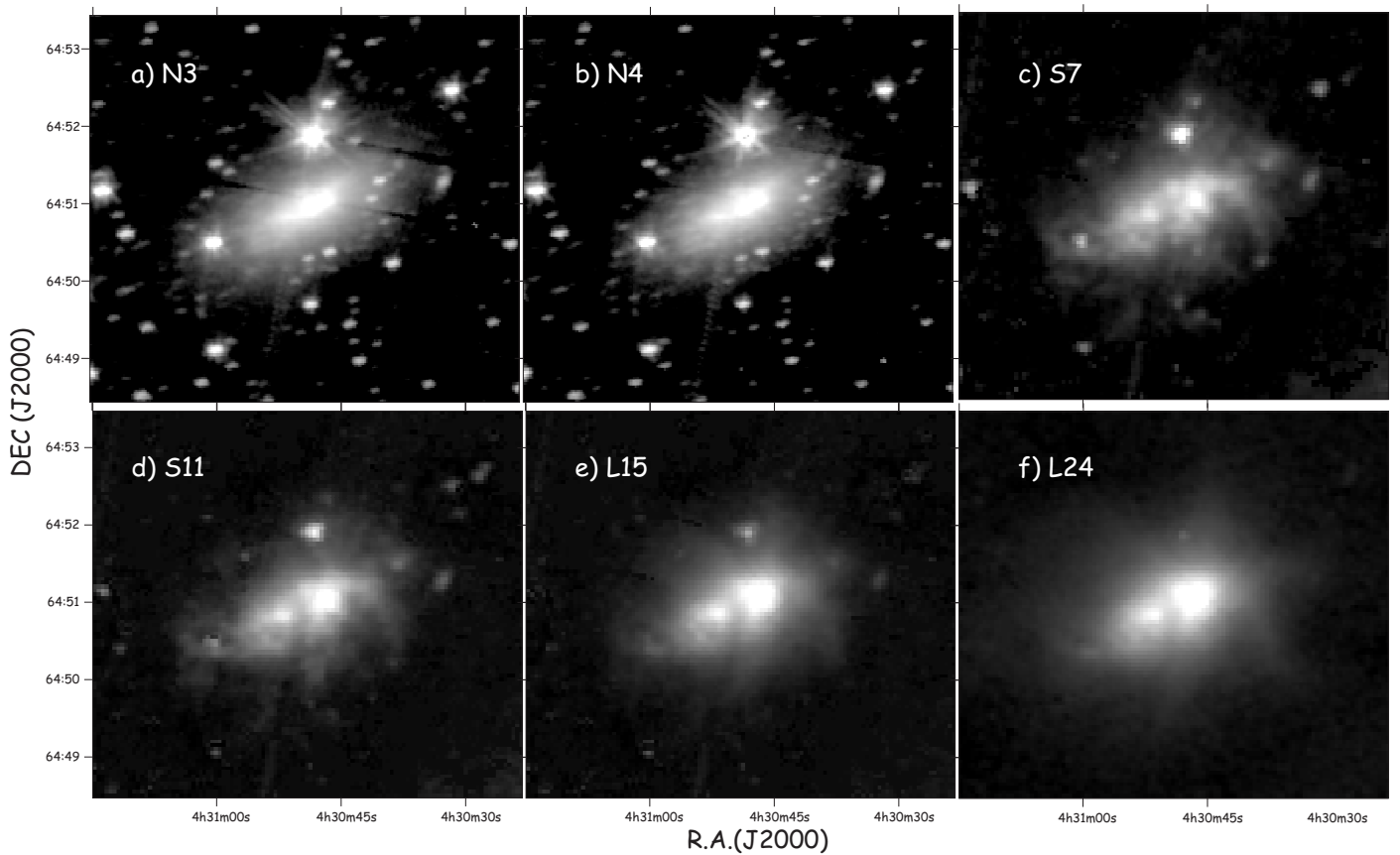}
      \caption{AKARI/IRC 6-band images of NGC1569.  The image size is 
      $5\farcm5 \times 5\farcm$0. a) N3 (3.2\,$\mu$m), b) N4 (4.1\,$\mu$m), 
      c) S7 (7.0\,$\mu$m), d) S11 (11.0\,$\mu$m), e) L15 (15.0\,$\mu$m),
      and f) L24 (24\,$\mu$m).
              }
         \label{fig:image}
   \end{figure*}

Figure~\ref{fig:image} shows IRC 6-band images of NGC1569.  Uniform sky
background was subtracted from all the images.  The N3 and N4
images exhibit similar structures, in which stellar contributions are dominant.
The appearance of the galaxy drastically changes from the S7 image.
The emission in the N3 and N4 images is dominated by the stellar radiation,
whereas the emission longer than the S7 arises dominantly 
from the interstellar matter
of the galaxy.  The S7 image
clearly resolves two peaks seen in ISOCAM observations \citep{madden06}
and indicates spotty structures in the galaxy disk.  
The western peak 
is located at the slightly western side of SSC A
\citep[e.g.,][]{hunter00}.  The hot spot or the region of velocity crowding, which is
seen in the \ion{H}{i} map and suggested as the impact location of the
infalling gas onto the galaxy disk \citep{muhle05}, is located further west
of the western MIR peak.
The S7 band consists of major MIR UIR band emission (6.2, 7.7, and 8.2\,$\mu$m
bands) and thus traces the UIR band emission efficiently
\citep{ishihara07}.  Both the 11.3\,$\mu$m UIR band and
continuum emission could make a significant contribution to the S11 image.
The continuum emission dominates in the L15 and L24 bands, which
traces regions active in star formation \citep[e.g.][]{onaka07a, draine07}.
NGC1569 contains a number of SSCs and thus the
L15 and L24 bands are probably dominated by the contribution of
the emission from active star-forming regions.  The disk emission in S7 is dominated
by the UIR emission.
A filamentary structure also extends from the western edge of the 
galaxy disk to the south in the S7 image.  
This filament is also detected in the S11 image.
In the L15 and L24 images, the disk emission becomes smoother, whereas
the diffuse emission appears to extend to the north-east direction. 
A faint trace of the filamentary structure can be discerned,
but it is not as clear as in the S7 and S11 images.

\begin{figure*}
   \centering
   \includegraphics[width=\textwidth]{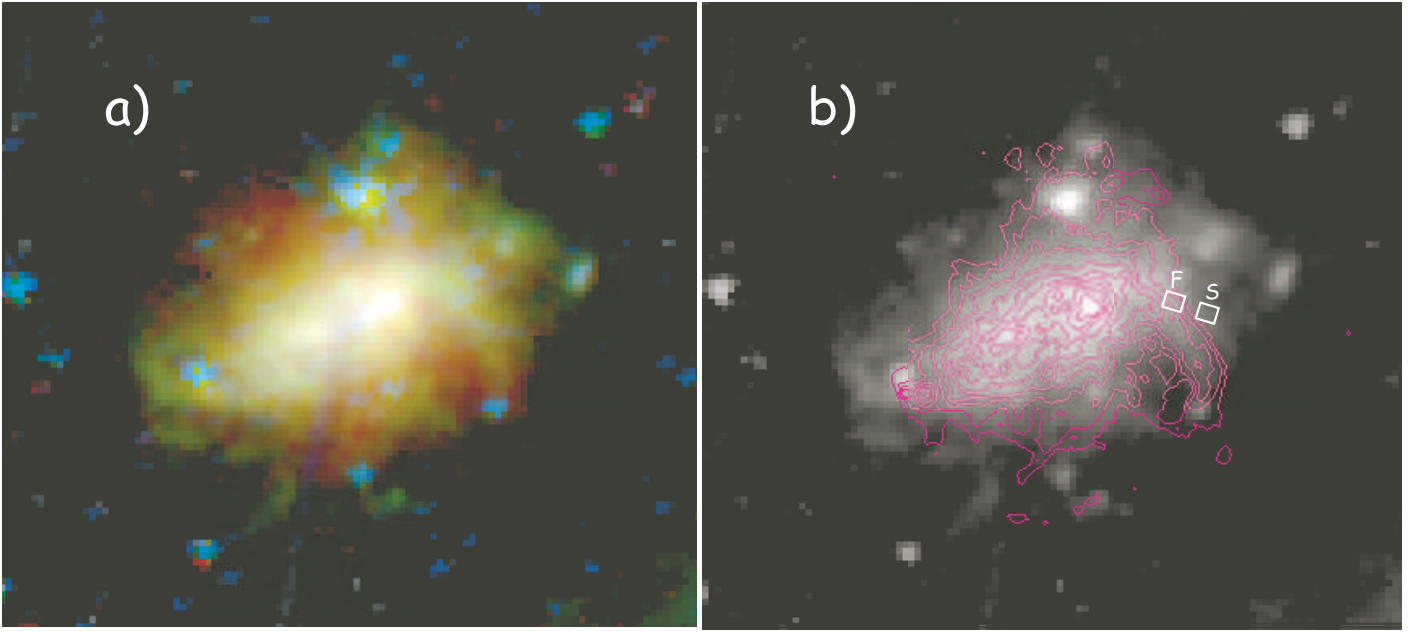}
      \caption{ a) Artificial 3-color image of NGC1569.  The N4 in blue,
      S7 in green, and L15 in red.  b) S7 image of NGC1569 superimposed with
      contours in a logarithmic scale of the H$\alpha$ emission 
      \citep{hunter04}.  The H$\alpha$ image \citep{hunter04} was taken from 
      the NASA/IPAC Extragalactic Database (NED).  The white boxes indicate
      the positions where the filament (F) and sky (S) spectra were extracted
      (see text).
              }
         \label{fig:color}
   \end{figure*}

Figure~\ref{fig:color}a
presents a three-band color image (N4, S7, and L15), which indicates that
the filamentary structure is indeed bright in the S7 band (green) and there is no
appreciable associated stellar component (blue).
This structure corresponds to the western arm or filament 6 recognized
in H$\alpha$ images \citep{hodge74, hunter93, west08}.
The associated X-ray emission strongly infers that the filament was formed by
the shock of a galactic outflow originating in the SSC A region of 
the disk \citep{martin98, martin02, greve02}.  
Figure~\ref{fig:color}b shows the S7 image superimposed on
the H$\alpha$ emission in contours \citep{hunter04}.  It shows 
good agreement between the S7 and H$\alpha$ emission of the filament 
in position and
suggests that the entire S7 emission may be more extended 
than the H$\alpha$ emission around the galaxy.
It can also be seen that the S7 emission of the
filament is located at a slightly western side of the H$\alpha$ emission.
Since the X-ray emission peaks at the eastern side of the H$\alpha$ 
filament \citep{martin02}, the peak of the emission is close in position
to that of the X-ray, H$\alpha$, and 7\,$\mu$m emission from east to west.

A similarity between the S7 and S11 images infers that the 11.3$\mu$m band
emission is also bright in the filament.  The present imaging observations
appear to detect both UIR bands in the H$\alpha$ filament and
an extended component around NGC1569.  The H$\alpha$ filament is
probably formed by the outflow associated with the X-ray emission,
suggesting that the UIR band emission is also enhanced by the outflow.
We carried out spectroscopic observations of the filament to confirm
the presence of the UIR bands.

\subsection{Spectroscopy} 
\begin{figure*}
   \centering
   \includegraphics[width=\textwidth]{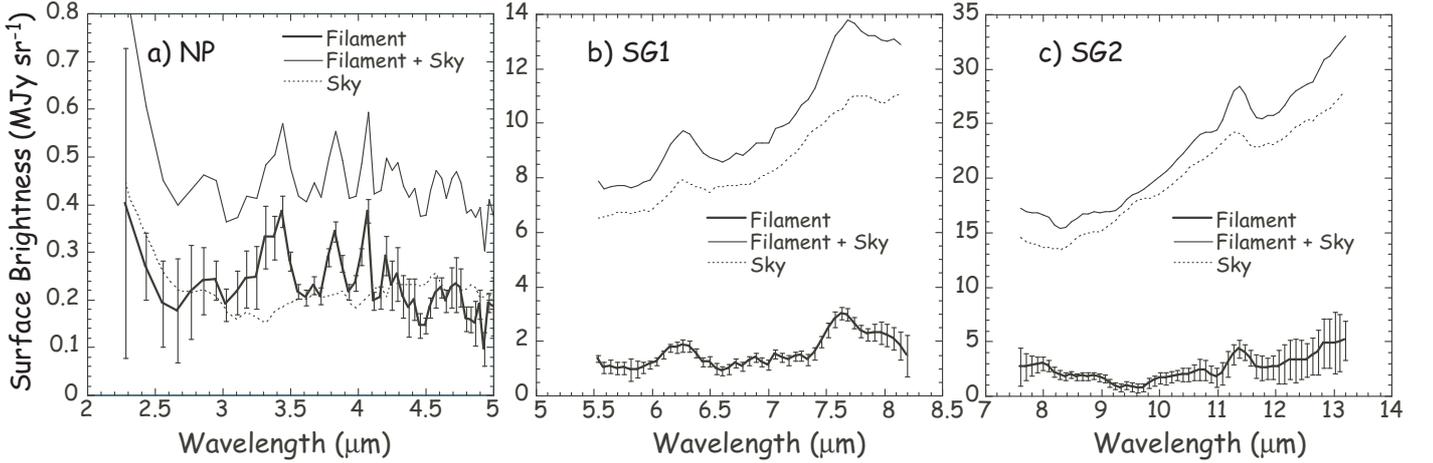}
      \caption{{\it AKARI}/IRC spectra of the H$\alpha$ filament of NGC1569: 
      a) NP, b) SG1, and c) SG2.  The thin solid lines indicate spectra
      at the filament position
      before subtraction of the sky background and the thin dotted lines
      show sky spectra.  The thick solid lines indicate filament spectra from
      which the sky spectra have been subtracted.  The error bars are indicated
      only in the subtracted spectra to avoid confusion.
              }
         \label{fig:spectrum}
   \end{figure*}

Figure~\ref{fig:spectrum} shows {\it AKARI}/IRC spectra taken at the
position of the H$\alpha$ filament.  The NP spectroscopy clearly shows
the continuum 
emission at the filament position over the background 
(Fig.~\ref{fig:spectrum}a).  It 
cannot be accounted for by photospheric emission from stars, which
should decrease far more sharply towards longer wavelengths.  
The NP spectrum also exhibits
the 3.3\,$\mu$m UIR band.  Two spiky features
at 3.75 and 4.05\,$\mu$m may be \ion{H}{i} recombination lines of Pf$\gamma$ and
Br$\alpha$, but the low spectral resolution of NP, even in 
the unsmoothed spectrum, makes
clear identification difficult.  If these are \ion{H}{i} recombination lines,
there should be a Pf$\delta$ line at 3.3\,$\mu$m that overlaps with the
UIR band.  Its intensity is
roughly $0.7 \times$ Pf$\gamma$ for the case B condition with the
electron temperature of $10^4$\,K and density of $10^4$\,cm$^{-3}$.  
Thus, even if the
contribution of Pf$\delta$ is taken into account, the presence of
the 3.3\,$\mu$m UIR band is secure.

The SG1 and SG2 spectra are dominated by the zodiacal emission, which
steeply increases towards longer wavelengths (Figs.~\ref{fig:spectrum}b and
\ref{fig:spectrum}c).  
The level of
the sky background emission is in good agreement with the estimated
zodiacal light intensity from
COBE observations \citep{kelsall98}.  
The sky-subtracted SG1 and SG2 spectra do not show a steep rise
towards longer wavelengths, implying that star formation is not
very active in the filament \citep{onaka07a}.
The sky spectra also appear to contain
UIR bands, which can be attributed partly to the 
foreground Galactic emission.  There may also be a contribution
from the emission of NGC1569 (see Fig.~\ref{fig:color}b).  
Thus the sky-subtracted filament spectra may underestimate the
emission from the filament.
According
to the investigation by \citet{sakon04}, the contribution of the
Galactic emission to the UIR bands is estimated approximately 
from the FIR intensity about the position of NGC1569.  
The contribution to the observed
UIR band emission is estimated to be about 30--70\%
from the FIR intensity of 
40 and 70 MJy sr$^{-1}$ at 100 and 140\,$\mu$m, respectively.
This number is compatible with the UIR band emission in the sky spectrum.  
The background-subtracted spectra clearly exhibits
the UIR bands at 6.2, 7.7, and 11.3\,$\mu$m
in the filament, confirming that the structures seen
in the S7 and S11 images are dominated by the UIR band emission.
The 8.6\,$\mu$m UIR band is faint and is not detected by the
present spectra.

\section{Discussion}
\subsection{UIR bands in the filament}
The present observations detect the UIR bands 
associated with a H$\alpha$ filament in NGC1569.  
Both FIR and UIR band emission has been detected in outflows or halo regions 
of several galaxies.  Very extended UIR band emission has been detected
in the outflow from M82 \citep{engelbracht06,kaneda10}.  
The 3.3\,$\mu$m UIR band emission associated with the outflow 
has been detected in NGC253 \citep{tacconi05}, in which
emission further away from the galaxy has been detected in the
FIR \citep{kaneda09b}.
The UIR band emission from the galactic halo has also been detected
for several galaxies based on ISOCAM observations 
\citep{irwin06, irwin07, galliano08b}, which indicates that the
band ratio of either 6.2 or 7.7\,$\mu$m to the 11.3\,$\mu$m band emission
becomes lower in the halo than in the disk region.  

   \begin{figure*}
   \centering
   \includegraphics[width=\textwidth]{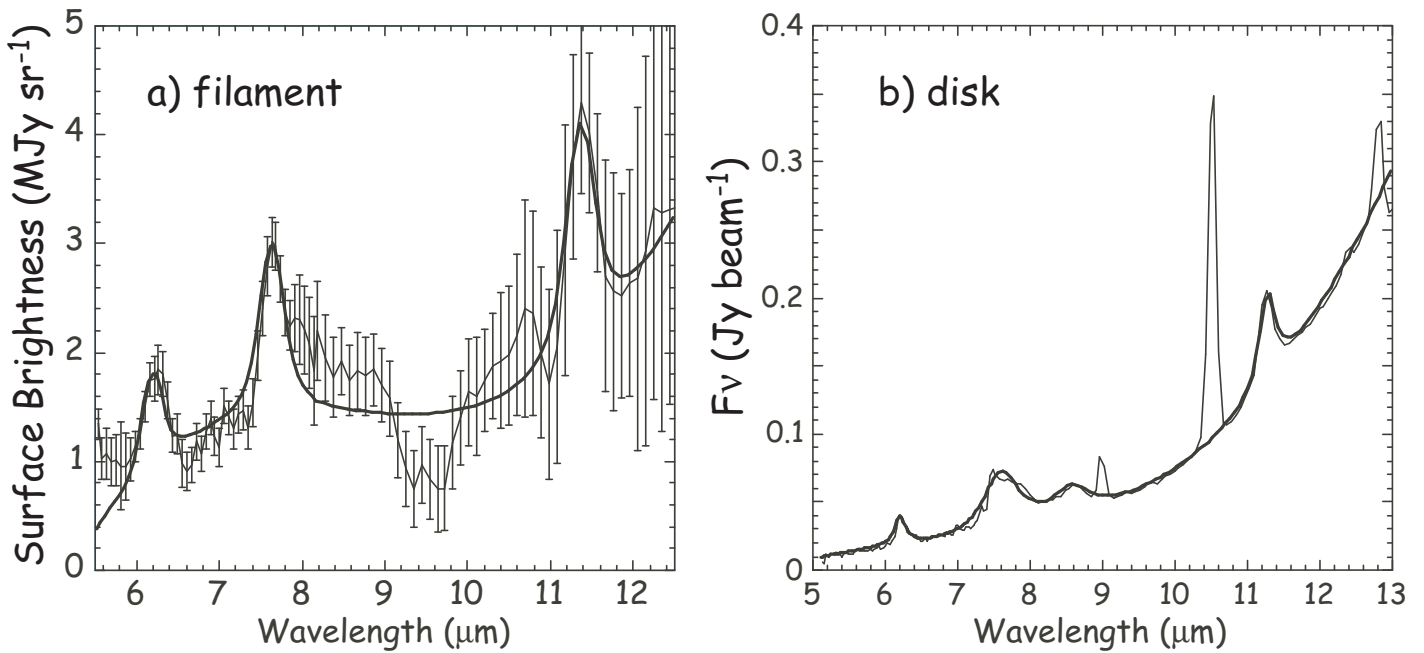}
      \caption{a) IRC spectrum of the filament of NGC1569 (thin line) and
      the fitted line (thick line).  b) IRS spectrum of the disk of
      NGC1569 (thin line) and the fitted line (thick line).  Sharp lines
      not fitted are ionic forbidden lines ([\ion{Ar}{III}]\,9.0\,$\mu$m,
      [\ion{S}{IV}]\,10.5\,$\mu$m, and [\ion{Ne}{II}]\,12.8\,$\mu$m).
              }
         \label{fig:UIR}
   \end{figure*}
   
To investigate the UIR band ratio in the filament of NGC1569,
the spectrum is fitted with a combination of a continuum emission and
Lorentzians given by

\begin{equation}
 F_\nu(\lambda) = \sum_{i=0}^{n} a_i \lambda^i 
                  + \sum_{i=1}^{m} \frac{h_i}{1+(\lambda_i/\Delta\lambda_i)^2
                   (1/\lambda - 1/\lambda_i)^2} \,, 
\label{eq:fit}
\end{equation}
where $\lambda$ indicates the wavelength, and
the first and second terms approximate the continuum and 
the UIR band emission, respectively.  The parameters $n$ and $m$
are set to be 3: only three UIR bands (6.2, 7.7, and 11.3\,$\mu$m)
are taken into account in the fit since the 8.6\,$\mu$m is weak
and its parameters cannot be constrained well by the fit.  
The parameters $\lambda_i$,
$\Delta\lambda_i$, and $h_i$ represent the central wavelength, 
width, and strength of each band component.
The fitted curve is shown in Fig.~\ref{fig:UIR}a together with the
observed spectrum.  Equation~(\ref{eq:fit}) provides a fairly close fit
except for a slight underestimate around 8--9\,$\mu$m and an overestimate
in 9--10\,$\mu$m.   The former could be attributed to the neglect of
the 8.6\,$\mu$m band.  A longer wavelength component of the 7.7\,$\mu$m
band may also be a contributor (see below).  The latter may be possibly attributed
silicate absorption.

The band strength $S_i$ is calculated by integrating the
Lorentzian in the wavenumber space and is given by

\begin{equation}
 S_i = \pi h_i \Delta\lambda/\lambda_i^2 \,.
\end{equation}
Table~\ref{tab:UIR} summarizes the band strength relative to the
11.3\,$\mu$m band.

   \begin{table}[!ht]
      \caption[]{UIR band strength ratio relative to the 11.3\,$\mu$m band}
         \label{tab:UIR} 
       \centering
      \begin{tabular}{l c c }
      \hline
      \hline
 	Band & Filament & Disk   \\
      \hline
        6.2\,$\mu$m  & $0.94\pm0.29$ & $0.71\pm0.08$ \\
        7.7\,$\mu$m  & $1.45\pm0.31$ & $3.54\pm0.21$ \\
        8.6\,$\mu$m  & ...$^{\mathrm{a}}$ & $1.37\pm0.13$ \\
      \hline
      \end{tabular}
      \begin{list}{}{}
      \item[$^{\mathrm{a}}$] The 8.6\,$\mu$m band is not included in the fit.
      \end{list}
    \end{table}

To compare with the UIR bands in the disk, spectra 
of the NGC1569 disk taken with the Infrared Spectrograph (IRS) 
on {\it Spitzer} 
in the low resolution modules were retrieved from the archival database
\citep[AOR key 9001984;][]{tajiri08}.  
The IRS spectra taken at the position near SSC A
(R.A: 04$^\mathrm{h}$ 30$^\mathrm{m}$ 48$\fs$17 Dec: 
+64\degr 50\arcmin 54\farcs2) were extracted.  The IRS spectrum is fit with 
Eq.~(\ref{eq:fit})
in the same manner as the filament spectrum except that the 8.6\,$\mu$m band
is included.  The IRS spectrum and the fitted spectrum
are also shown in Fig.~\ref{fig:UIR} and the relative band ratios are
summarized in Table~\ref{tab:UIR}.  The relative strength of the
6.2\,$\mu$m band in the filament seems to be slightly greater than
in the disk.  However, the 6.2\,$\mu$m band strength of the filament
has a large uncertainty and
it is difficult to draw conclusions about the comparison with
the disk spectrum.  The relative strength of the 7.7\,$\mu$m band is
clearly weaker
in the filament than in the galaxy disk.  This is the same trend
as derived from photometric measurements 
of halo regions of other galaxies \citep{irwin06, galliano08b}.
The 7.7 to 11.3\,$\mu$m band ratio of the IRS spectrum
is in the range of values measured for normal and starburst galaxies,
while the ratio for the filament is similar to those of galaxies with weak AGNs
or elliptical galaxies \citep{smithjd07, onaka08}.  

An additional inspection suggests that the weak strength of the 7.7\,$\mu$m
band in the NGC1569 filament may be attributed to its narrow band
width relative to those in the disk spectrum ($\Delta\lambda = 0.13\pm0.04$\,$\mu$m to 
$0.31\pm0.02$\,$\mu$m).  The narrow width is also suggested in 
Fig.~\ref{fig:UIR}.  The 7.7\,$\mu$m UIR band is known to consist of
more than two components \citep{peeters02}.  The filament spectrum
suggests that the longer wavelength component of the 7.7\,$\mu$m 
band may be weak or absent.  There may be excess above the fitted curve
around 7.8--7.9\,$\mu$m, which is not included in the band 
strength estimate.
This excess increases the band strength only within its uncertainty
and does not change our conclusion that the total strength of the
7.7\,$\mu$m band is weak in the filament.  The present observation
suggests that the low band strength of the 7.7\,$\mu$m band in the filament may
be attributed to a change in one of the components.  

A low ratio of the 7.7 to 11.3\,$\mu$m band
is also suggested in the outer region of our Galaxy \citep{sakon04}
as well as in the interarm region of NGC6946 \citep{sakon07}.
Elliptical galaxies are an extreme case for which the 6.2 and 7.7\,$\mu$m
bands are almost absent \citep{kaneda05, kaneda08}.  \citet{kaneda08}
attributed the abnormal UIR band strengths in elliptical galaxies to
the dominance of neutral PAHs.  The weak 6.2 and 7.7\,$\mu$m bands
relative to the 11.3\,$\mu$m band tend to be seen in tenuous regions
and may have a common cause or be the result of processing of the band carriers
in these environments \citep[e.g.,][]{onaka08}.

\citet{kaneda09b} suggest that the FIR emission detected at 6--9\,kpc from
the galaxy disk of NGC253 may come from the emission of outflowing dust
entrained by superwinds.  The UIR band carriers in the filament of NGC1569
may also be entrained by the outflow of NGC1569, which formed the H$\alpha$
filament.  The velocity of the filament is derived to be about 
90\,km\,s$^{-1}$ \citep{west08}.
If the filament were produced by the outflow produced by
the star-formation activity of SSC A, the distance 
of 490\,pc from SSC A to the position of the IRC spectrum
would infer an expansion timescale of about 5.3 Myr.  
Observations with {\it Chandra} suggest that the electron temperature and
density of the bubble are $3.51 \times 10^6$\,K and
$0.035$\,cm$^{-3}$, respectively \citep{ott05}.  
\citet{jones96} show that thermal sputtering 
efficiently destroys dust grains in fast shocks ($ \le 200$\,km\,s$^{-1}$)
and that the thermal sputtering yield depends on the electron temperature
and density, which can be applied even to very small grains.
Using the equation given by \citet{tielens94}, the thermal sputtering timescale for
grains of 1\,nm in the bubble of NGC1569
is found to be $1.3 \times 10^3$ yr, which is much shorter than the expansion
timescale of the filament.  Thus it seems unlikely that
the band carriers in the filament originate in the galaxy disk and
are entrained by the outflow without destruction.  
It is also unlikely that AGB stars produce a large amount of
the band carriers since there appears to be neither appreciable
stellar components nor strong star-forming activities in the filament.
An alternative
possibility for the origin of the carriers may be the fragmentation from
large carbonaceous grains in shocks that produce H$\alpha$ emission
\citep{jones96}.  The presence of the 3.3\,$\mu$m band suggests that
the smallest band carriers exist in the filament, which may be
consistent with the fragmentation origin.
UIR band carriers formed from fragmentation may have different
properties from those in the galactic disk and exhibit the 
weaker 7.7\,$\mu$m band.

\subsection{NIR excess continuum}
The present observations indicate the presence of excess continuum emission
in the NIR (2.5--5\,$\mu$m) in the filament.  NIR excess continuum emission
was first reported in reflection nebulae \citep{sellgren83} and then found
in normal galaxies \citep{lu03}.  It is also seen 
in the diffuse Galactic emission
\citep{flagey06}.  \citet{sellgren84} suggests that the excess emission
in reflection nebulae may be attributed to stochastically heated
3-dimensional grains consisting of 45--100 carbon atoms.  The continuum
emission is also found to have a distinct spatial distribution from the 
3.3\,$\mu$m UIR band emission \citep{an03}.

Redder colors in the NIR have been measured for irregular/Sm galaxies
than for spirals \citep{pahre04, engelbracht05, smith07}.  They have been
attributed to younger stars \citep{pahre04}, hot dust \citep{engelbracht05,
hunter06}, nebular emission, or to the 3.3\,$\mu$m UIR band emission.
\citet{smith09} investigate the origin of excess emission at 4.5\,$\mu$m
in dwarf galaxies in detail.  They discuss the possibilities of a
contribution from Br$\alpha$, the reddening
of starlight, and a nebular continuum as the origin of the excess, 
concluding that a combination
of these three may account for the 4.5\,$\mu$m excess and no significant
contribution of hot dust is necessary, although hot dust emission cannot
be completely excluded.

\begin{figure}
   \centering
   \includegraphics[width=9cm]{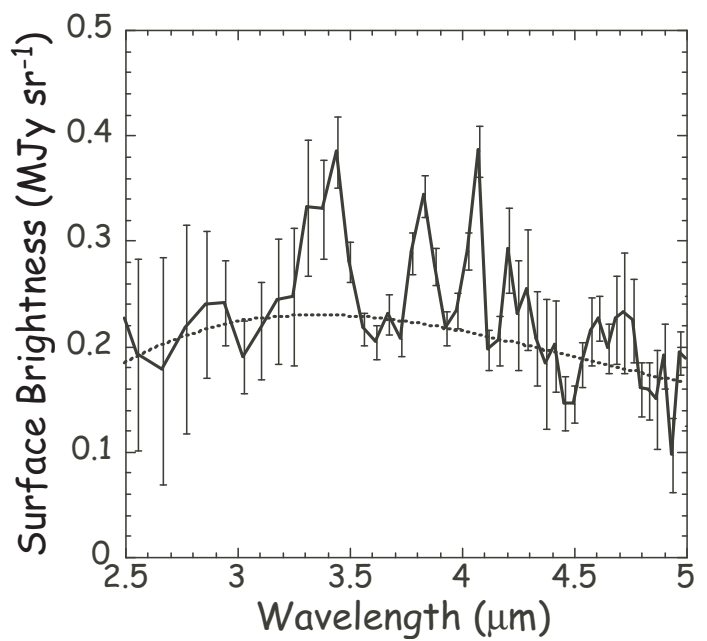}
      \caption{NP spectrum of the filament (solid line) together with
      the fitted modified black body (dotted line).  The fitted curve
      is a black body of $T=868$\,K with the emissivity of $\propto
      \lambda^{-2}$.
              }
         \label{fig:hot}
   \end{figure}

The NP spectrum clearly shows the presence of
excess continuum emission in addition to the 3.3\,$\mu$m UIR band and
line emission, if any.  The continuum spectrum is rather flat in the NIR spectral
range and is distinct from stellar photospheric emission.  
\citet{lu03} suggest that the excess emission seen in normal galaxies
is well fitted with a modified black body, which can be attributed to emission
from hot dust.
Figure~\ref{fig:hot} shows the NP spectrum of the filament together with 
a fitted modified black body of $T=868$\,K with the emissivity of $\propto
\lambda^{-2}$.  The modified black body curve closely reproduces the observed spectrum,
suggesting a similarity to the excess continuum in normal galaxies
and a hot dust origin for the excess emission. 
If the spiky features
at 3.75 and 4.05\,$\mu$m are assumed to be Pf$\gamma$ and Br$\alpha$, 
the associated
free-free emission is expected.  Simple estimates of the intensities of the two
lines obtained by integrating over the continuum infer $2.8 \times 10^{-9}$ and
$2.4 \times 10^{-9}$\,W\,m$^{-2}$\,sr$^{-1}$ for Pf$\gamma$ and Br$\alpha$,
respectively.  The line ratio differs significantly from the case A or B, which
infers that there is a large uncertainty in the estimated line intensities.  
Far stronger 
Pf$\beta$ than Pf$\gamma$ emission should be present around 4.65\,$\mu$m,
where only a marginal hump is seen.
The strength of the hump is compatible with the estimated  Br$\alpha$ 
intensity.  Thus we estimate the intensity of the free-free emission from
the Br$\alpha$ intensity.
For the case B with the electron temperature of $10^4$\,K,
the free-free intensity expected from the Br$\alpha$ intensity
is about 0.03 MJy\,sr$^{-1}$.  Therefore, the free-free emission alone cannot
account for the observed excess continuum emission
and a hot dust contribution may be needed even considering the large uncertainty
in the estimated line intensity.  

\section{Summary}
We have presented {\it AKARI}/IRC observations of the starburst dwarf galaxy
NGC1569.  Imaging observations have detected the MIR UIR band emission
in a H$\alpha$ filament, which is formed by the outflow originating
in the star-formation activities of the NGC1569 disk.  
Subsequent spectroscopic observations
also with the IRC clearly confirm the presence of the UIR bands at 3.3,
6.2, 7.7, and 11.3\,$\mu$m in the filament.  A comparison with the
IRS spectrum of the disk of NGC1569 suggests that the UIR bands in
the filament have a lower ratio of the 7.7 to 11.3\,$\mu$m band strength
than in the galaxy disk.
The low band ratio is also measured for the outflow or halo regions
of other galaxies in addition to elliptical galaxies and the interarm regions of 
our Galaxy.  Common characteristics may be attributed to
the processing of the band carriers in these tenuous regions.

Thermal sputtering in the outflow is estimated to be very efficient and
we propose that the band carriers cannot survive in the outflow.
Alternatively, we suggest that 
the band carriers may be formed by the fragmentation of
larger carbonaceous grains in the shock that produces the
H$\alpha$ emission.

The presence of excess continuum emission in the NIR is also indicated by the 
present spectroscopy in addition to the 3.3\,$\mu$m UIR band.
The present
spectrum implies both that the excess continuum emission cannot
be accounted for solely by the free-free emission estimated from the
Br$\alpha$ that is possibly detected and that hot dust emission
may be an important contributor.

\begin{acknowledgements}
This work is based on observations with {\it AKARI}, a JAXA project
with the participation of ESA.  The authors thank all the members of
the {\it AKARI} project and the members of the Interstellar and Nearby
Galaxy team for their help and continuous encouragements.  Part of
this work is based on observations made with the {\it Spitzer Space
Telescope}, which is operated by the Jet Propulsion Laboratory (JPL),
California Institute of Technology under a contract with the 
National Aeronautics and Space Administration (NASA).
This work has also made use of the NASA Extragalactic Database (NED)
which is operated by the JPL, California Institute of Technology under
contract with NASA.
The authors thank T. L. Roellig and Y. Y. Tajiri for providing us with the IRS
spectrum of NGC1569.  They also thank S. M\"uhle for providing
their \ion{H}{i} data.    This work is supported by a Grant-in-Aid
for Scientific Research from the Japan Society for the Promotion of Science
(no. 18204014).
\end{acknowledgements}

\end{document}